\documentclass[preprint,12pt]{elsarticle}

\usepackage{multicol}
\usepackage{graphicx}
\usepackage{booktabs}
\usepackage{amssymb,bm,mathrsfs,bbm,amscd}
\usepackage[tbtags]{amsmath}
\usepackage{lastpage}
\usepackage{verbatim}
\usepackage{comment}

\journal{}

\begin{document}

\begin{frontmatter}



\title{Antineutrino spectral anomaly within the neutron spectrum\tnoteref{label1}}
 \tnotetext[label1]{Corresponding author}
\author{Xubo Ma\corref{cor1}\fnref{label2}}
 \ead{maxb@ncepu.edu.cn}
\author[label2]{Le Yang}

\address[label2]{School of Nuclear Science and Engineering, North China Electric Power University, Beijing 102206, China \fnref{label2}}

\begin{abstract}
Recently, three antineutrino experiments at Daya Bay, Double Chooz, and RENO measured the neutrino mixing angle $\theta_{13}$, each using a nuclear reactor power plant. However, significant discrepancies were found, both in the absolute flux and spectral shape. In the reactor, the neutrons have a range of energies, with different neutron energies generating different fission yields. The different fission yields may be the reason for the antineutrino spectrum discrepancies. In our study, the neutron spectrum has been analyzed to understand the possible reasons for the discrepancies. In comparing results from the Huber--Muller model, we found that a spectral bump appears in the energy region from 5--7~MeV. Nevertheless, the differences between the average antineutrino spectrum and the only-thermal antineutrino spectrum are small. That is, they are unable to account for this bump. However, in the energy region of 7--8~MeV, the neutron flux induces a decrease in the antineutron flux, and therefore, the distribution of the antineutrino spectrum is affected, especially in the high-energy region.

\end{abstract}


\end{frontmatter}

\section{Introduction}

Recently, three antineutrino experiments at Daya Bay\cite{Dayabay}, Double Chooz\cite{DoubleChooz}, and RENO\cite{RENO} measured the neutrino mixing angle $\theta_{13}$. However, significant discrepancies were found, both in the absolute flux and spectral shape. The discrepancies in the absolute flux were collectively called the "reactor neutrino anomaly", which first appeared in print by Mention and colleagues\cite{mention}. For the antineutrino spectral shape, a 2.9$\sigma$ deviation was found in the measured positron energy spectrum of the inverse beta decay compared with predictions. Specifically, an excess of events at energies of 4--6~MeV was found in the measured spectrum \cite{spc1Dayabay, spc2Dayabay,spc2Chooz}, with a local significance of 4.4$\sigma$. These results have made it obvious that neutrino fluxes are not as well understood as had been thought. At present, the physical processes giving rise to the neutrino spectral bump is unclear. Much effort has been spent on the reactor antineutrino anomaly, which arose from improved calculations of the antineutrino spectra derived from a combination of information from nuclear databases referencing $\beta$ spectra\cite{a7,a8,a9,a10}.

In essence, two approaches were applied to estimate the antineutrino spectra. One is called the "$\beta^{-}$ conversion". The energy spectra of the $\beta^{-}$ from beta decay were measured to estimate the corresponding $\bar{\nu}_{e}$ emission for the fissile isotopes $^{235}$U, $^{239}$Pu, and $^{241}$Pu from ILL reactor in the 1980s\cite{a8,a9,a10}. More recently, a similar measurement was made for $^{238}$U\cite{u238exp}. For a single measured $\beta^{-}$ decay spectrum, the corresponding $\bar{\nu}_{e}$ spectrum can be predicted with high precision. The other method is called the "\emph{ab initio} method". Using the measured $\beta^{-}$ decay parameters and fission yields of the nuclear library, the antineutrino spectra of each isotope can be evaluated by summing each measured $\beta^{-}$ spectra of the fission daughters\cite{fallot,hayes,dwyer}. This introduces uncertainties of a few percent in the corresponding predictions of such calculations. The antineutrino spectrum and flux vary with fuel burnup, and a recent study on the evolution of the reactor antineutrino flux and spectrum at Daya Bay\cite{fluxevo} has shown that a 7.8\% discrepancy between the observed and predicted $^{235}$U yields suggests that this isotope may be the primary contributor to the reactor antineutrino anomaly. The impact of fission neutron energy on reactor antineutrino spectra were also studied in \cite{neutronspec}, but no spectral effect concerning the spectral bump is given. In this study, the fission fraction variation with burnup was taken into account and a calculation method of the average cumulative fission yield was proposed to calculate the antineutrino spectrum. We discuss a calculation of the $\bar{\nu}_{e}$ spectrum from the nuclear JENDL3.1 database. Besides the nuclear database, the fission yield of each isotope is also needed. The fission yield data are taken from ENDF/B-VII.1 library. The \emph{ab initio} approach was applied and about 1100 isotopes nuclear data were used.

In Sec.~\ref{neutronpsc}, we describe the reactor simulation used in the study and present the fission yield difference relevant to the fission neutron energy. In Sec.~\ref{antispc}, we describe how to calculate the antineutrino spectrum using the nuclear databases. In Sec.~\ref{averagecu}, the method to calculate the average cumulative fission yield is proposed taking into account the fast neutron fission fraction contribution. An analysis of the antineutrino spectrum ratio obtained from the Daya Bay experiment measurements by taking into account fast neutron fission was performed, and no spectral bump was found. However, when compared with that of the Huber--Mueller model, a bump at 5--7~MeV was found. The differences in the antineutrino spectrum comparted with the neutron spectrum are small, and therefore neutron spectrum cannot be the main reason for the bump. These results are discussed in Sec.~\ref{bumpan}. The last section presents a summary.

\section{Neutron spectrum}
\label{neutronpsc}
\begin{figure}
\begin{center}
\includegraphics[width=9cm]{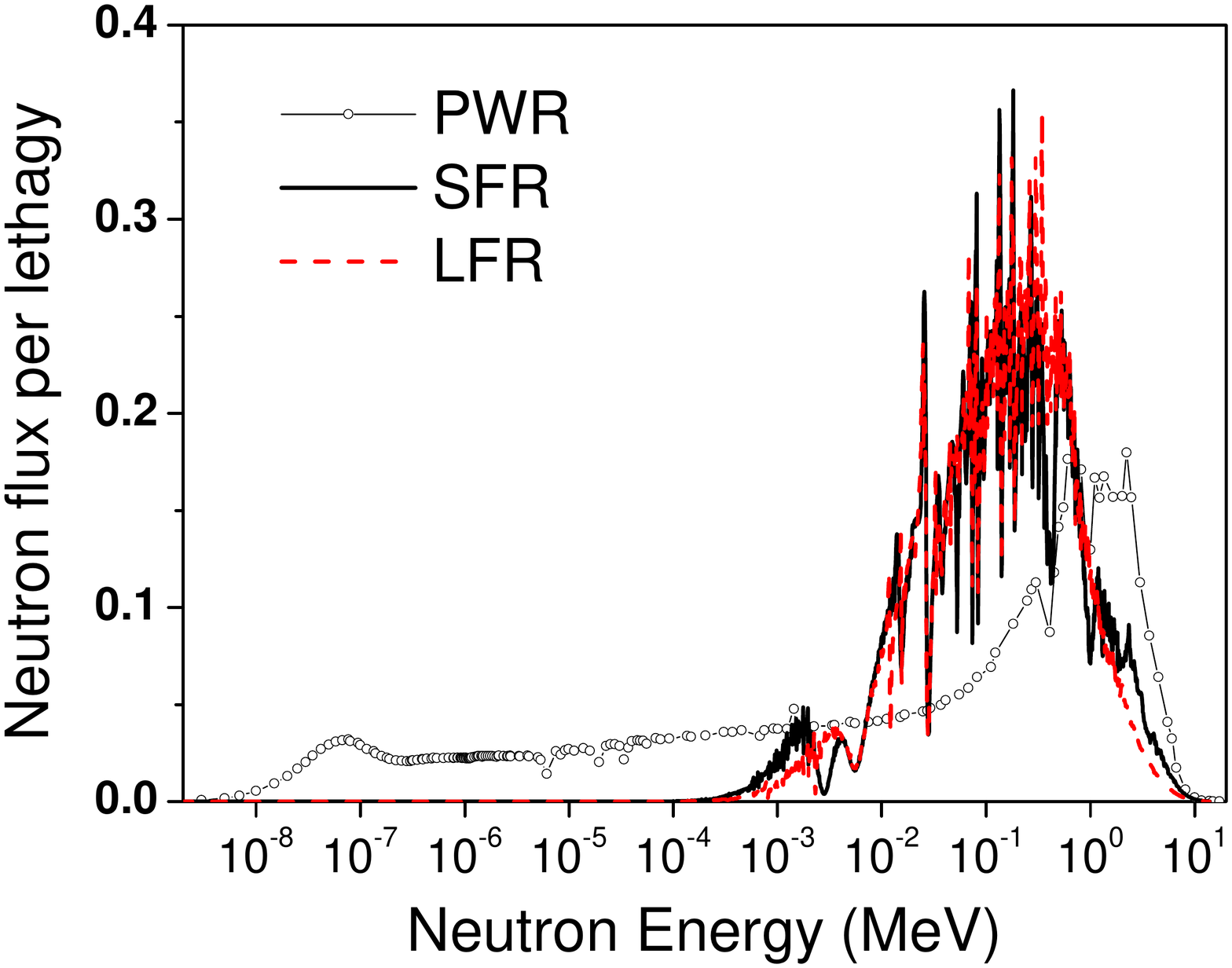}
\caption{Neutron spectra for pressure water reactor (PWR), sodium fast reactor (SFR), and lead-cooled fast reactor (LFR)}
\label{spectraneu}
\end{center}
\end{figure}

Most of the reactor neutrino experiments are using a pressurized water reactor (PWR) as a neutrino source. The neutron energy distribution in the reactor is called the neutron spectrum, and different types of reactors have different neutron spectra. The neutron spectra of a PWR, a sodium fast reactor, and a lead-cooled fast reactor are shown in Fig.~\ref{spectraneu}.

Most neutrons are observed to have high energies in the sodium and lead-cooled fast reactors. However, there are also many fast neutrons in the PWR that come from fission. The fission in the PWR core mainly comes from four isotopes, $^{235}$U, $^{238}$U, $^{239}$Pu, and $^{241}$Pu, and the total fission fraction of each of the four isotopes is more than 99\%. Typically, the total fission fractions of $^{235}$U and $^{239}$Pu are more than 80\%.

\begin{figure}
\begin{center}
\includegraphics[width=9cm]{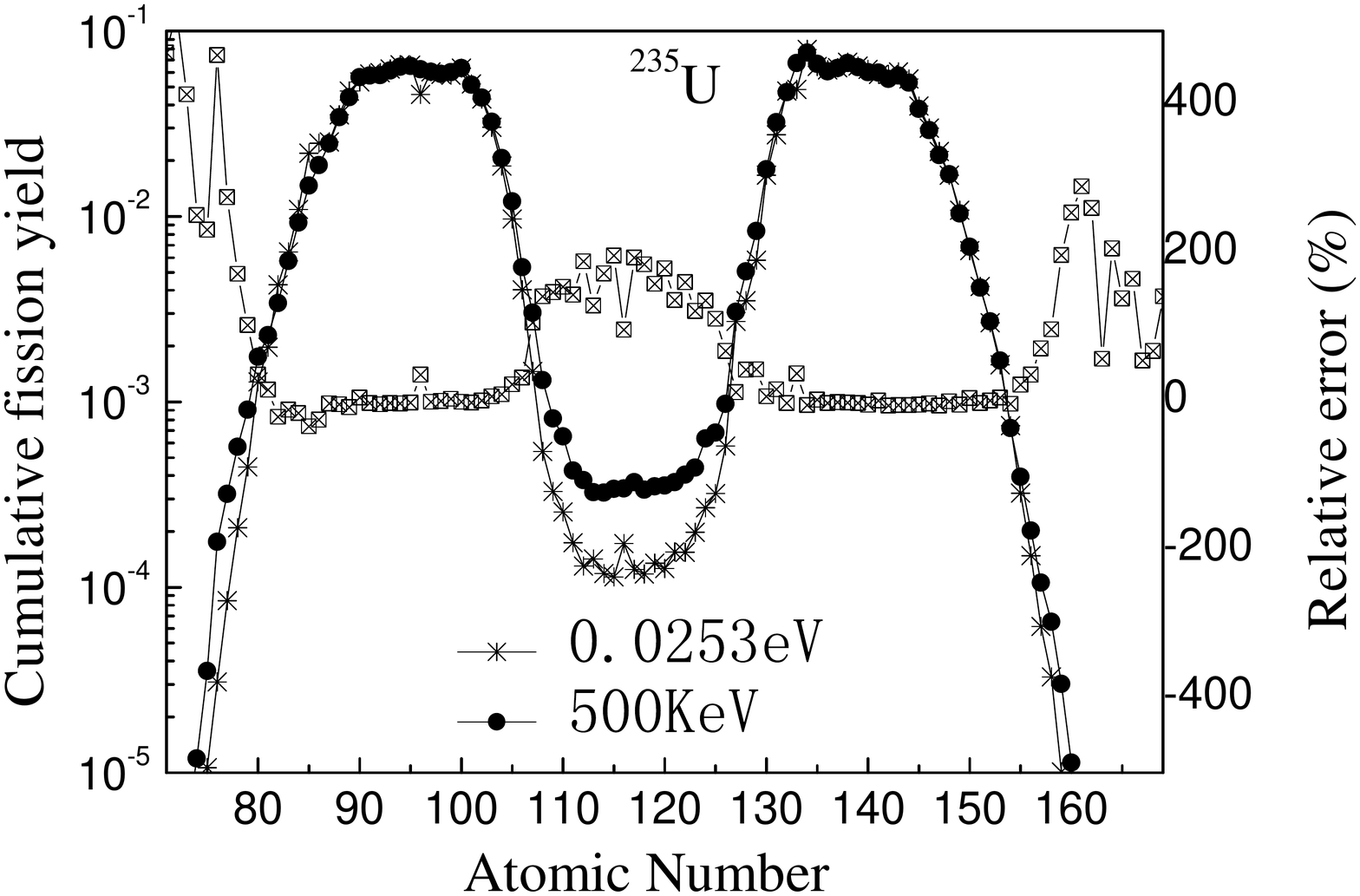}
\caption{Comparing cumulative fission yield differences between thermal and fast neutrons for $^{235}$U}
\label{yield235}
\end{center}
\end{figure}

\begin{figure}
\begin{center}
\includegraphics[width=9cm]{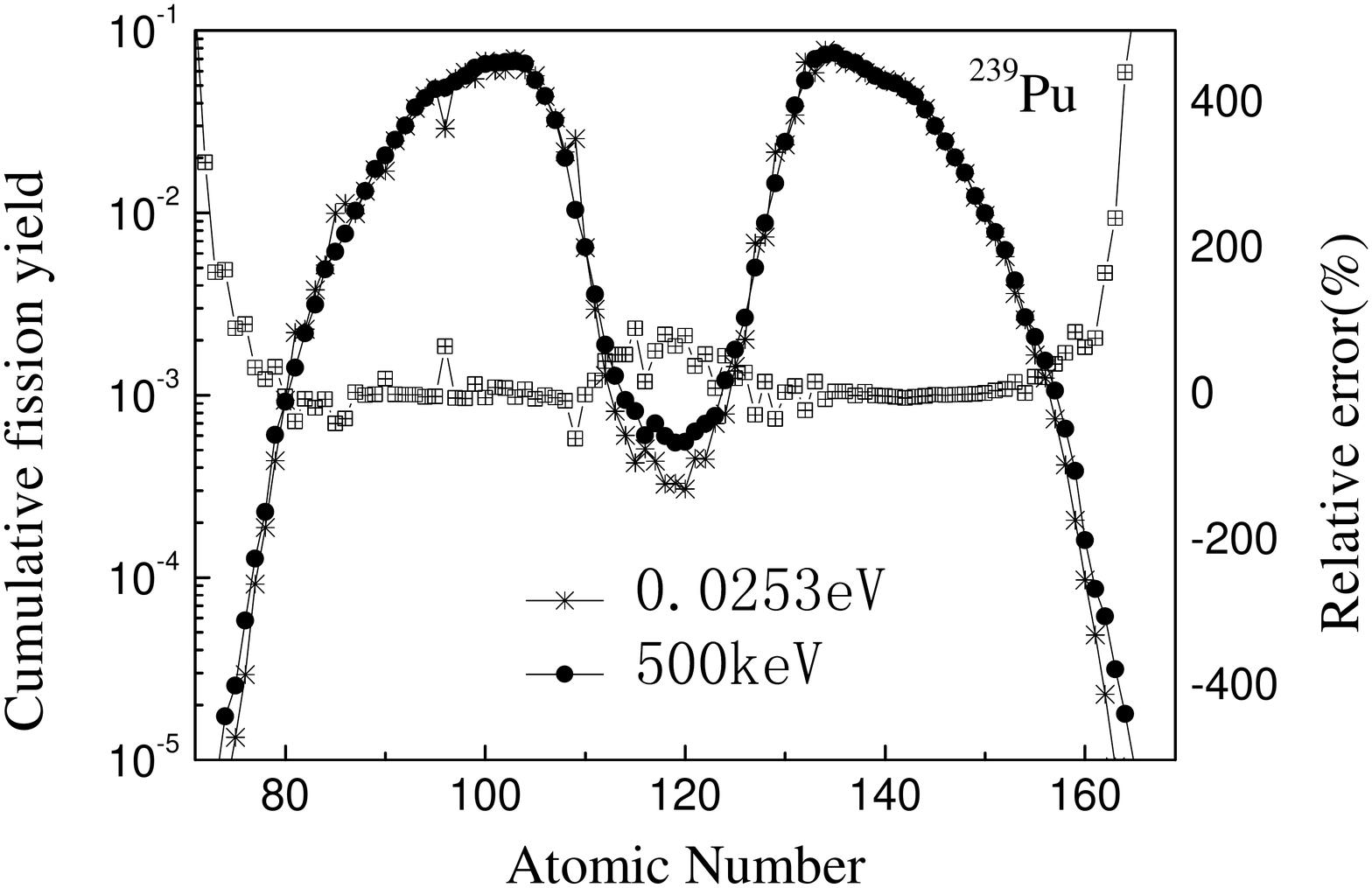}
\caption{Comparing cumulative fission yield differences between thermal and fast neutrons for $^{239}$Pu }
\label{yield239}
\end{center}
\end{figure}

The fission yields are different because of the different neutron-energy-induced fission reactions. The differences in thermal and fast neutrons for $^{235}$U and $^{239}$Pu are shown in Figs.~\ref{yield235} and \ref{yield239}. These differences appear at atomic numbers from 105 to 130, and the trend for $^{235}$U is the same as $^{239}$Pu.

In regard to the Daya Bay neutrino experiment, the power plant is a 990-MW$_{e}$ (electric power) reactor designed with three cooling systems (Framatome Advanced Nuclear Power, France). There are 157 fuel assemblies in the reactor core, each assembly having a 17$\times$17 design for a total of 289 fuel rods. The enrichment of $^{235}$U in the fuel for a one-and-a-half year refueling cycle is 4.45\%. They are all typical PWRs.


\section{Calculation of isotope antineutrino spectrum}
\label{antispc}
The \emph{ab initio} method for calculating the isotope antineutrino spectra are presented in [13,16,17]. For a system in equilibrium, the total antineutrino spectrum is given by
\begin{equation}
S(E_{\bar{\nu}})=\sum_{i=1}^{N}FP_{i}
\sum_{j=1}^{M} f_{ij}S_{ij}(E_{\bar{\nu}})
\label{totalspc}
\end{equation}
where $FP_{i}$ is the cumulative fission yield, $f_{ij}$ the branching ratio to the daughter level with energy $E_{e}$, and $S_{ij}(E_{\bar{\nu}})$ the antineutrino spectrum for a single transition with endpoint energy $E_{\bar{\nu}}=E_{0}-E_{e}$. The beta-decay spectrum $S_{ij}(E_{\bar{\nu}})$ for a single transition in the nucleus (Z,A) with end-point energy $E_{0}$ is
\begin{equation}
S(E_{e},Z,A)=S_{0}(E_{e})F(E_{e},Z,A)C(E_{e})[1+\delta(E_{e},Z,A)],
\label{singlespc}
\end{equation}
where $S_{0}(E_{e})=G_{F}^{2}p_{e}E_{e}(E_{0}-E_{e})/2\pi^{3}$, $E_{e}(p_{e})$ is the electron total energy (momentum), $F(E_{e},Z,A)$ the Fermi function needed to account for the Coulomb interaction of the outgoing electron with the charge of the daughter nucleus, and $C(E_{e})$ a shape factor for the forbidden transitions arising from additional electron momentum terms. The term $\delta(E_{e}, Z, A)$ represents a fractional correction to the spectrum. The primary corrections to the beta-decay analysis are radiative $\delta_{rad}$, finite size $\delta_{FS}$, and weak magnetism $\delta_{WM}$. For the allowed transitions $C(E_{e})$ = 1, the radiative, finite size, and weak magnetism corrections are taken from \cite{huber}. For the forbidden transitions, the shape factor and fractional weak magnetism corrections are taken from \cite{hayes}. In the present work, the cumulative fission yields are taken from the ENDF/B-VII.1 library\cite{endfyield}. The decay data are taken from a version of JENDL-3.1\cite{jendldecay}.

\section{Average cumulative fission yield calculation method}
\label{averagecu}
Fission products yield libraries based on evaluated nuclear data in the endf-6 format may be compiled for particular regions of a core with a local neutron spectrum or an entire core with an average neutron spectrum. Data on the yield of the $i$th fission products $FP_{i}(E_{k})$ for a prescribed neutron energy range $E_{k}$ are chosen for each actinide from a base set. The total number of nuclides---fission products in the initial, standard, nuclear data files---ranges from 800 to 1700. In ENDF/B, the fission-product yield is usually presented for the thermal point $E_{1}=$2.53 $\times$10$^{-8}$~MeV, the average point of the neutron fission spectrum $E_{2}$=0.4~MeV or $E_{2}$=0.5~MeV, and the energy $E_{max}$ $\sim$ 14~MeV. This three-group representation gives a non-unique representation of the yield in a neutron energy range and is important for the antineutrino spectrum calculation\cite{mitenkova}. In this study, the average cumulative fission-product yield in prescribed energy intervals of the spectrum is calculated for a selected nuclide, specifically,
  \begin{equation}
<FP_{i}>=\sum_{k} C_{k}FP_{i}(E_{k}),
  \label{eq1}
  \end{equation}
where the coefficients $C_{k}$ reflect the distribution of the number of fissions of a prescribed nuclide in the spectrum of the region under study. Here,
  \begin{equation}
C_{thermal}=\frac{\int_{0}^{E_{1}}dE\phi(E)\sigma_{f}(E)}
{\int_{0}^{E_{max}}dE\phi(E)\sigma_{f}(E)}
  \label{eq2}
  \end{equation}
  \begin{equation}
C_{fast}=\frac{\int_{E_{1}}^{E_{max}}dE\phi(E)\sigma_{f}(E)}
{\int_{0}^{E_{max}}dE\phi(E)\sigma_{f}(E)}
  \label{eq3}
  \end{equation}
where $E_{1}$ is the boundary energy between a thermal group neutron and a fast group neutron. From Eqs.~(\ref{eq2}) and (\ref{eq3}), the average fission-product yield, Eq.~(\ref{eq1}), is written
  \begin{equation}
<FP_{i}>=C_{thermal}FP_{i}(E_{1})+C_{fast}FP_{i}(E_{2}).
  \label{eq4}
  \end{equation}

\begin{figure}
\begin{center}
\includegraphics[width=9cm]{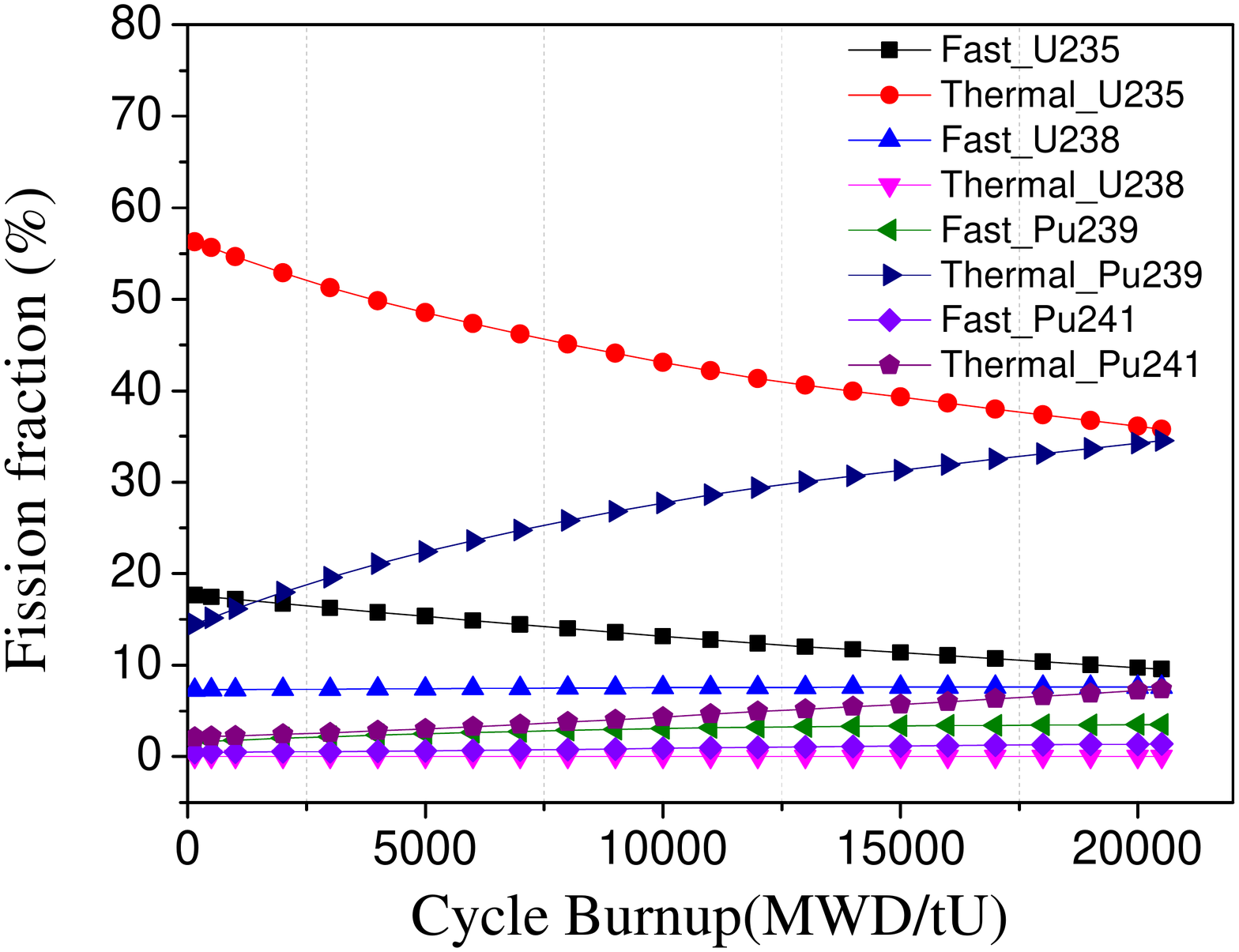}
\caption{Thermal and fast neutron fission fraction vs cycle burnup}
\label{thermalfastff}
\end{center}
\end{figure}

\begin{figure}
\begin{center}
\includegraphics[width=9cm]{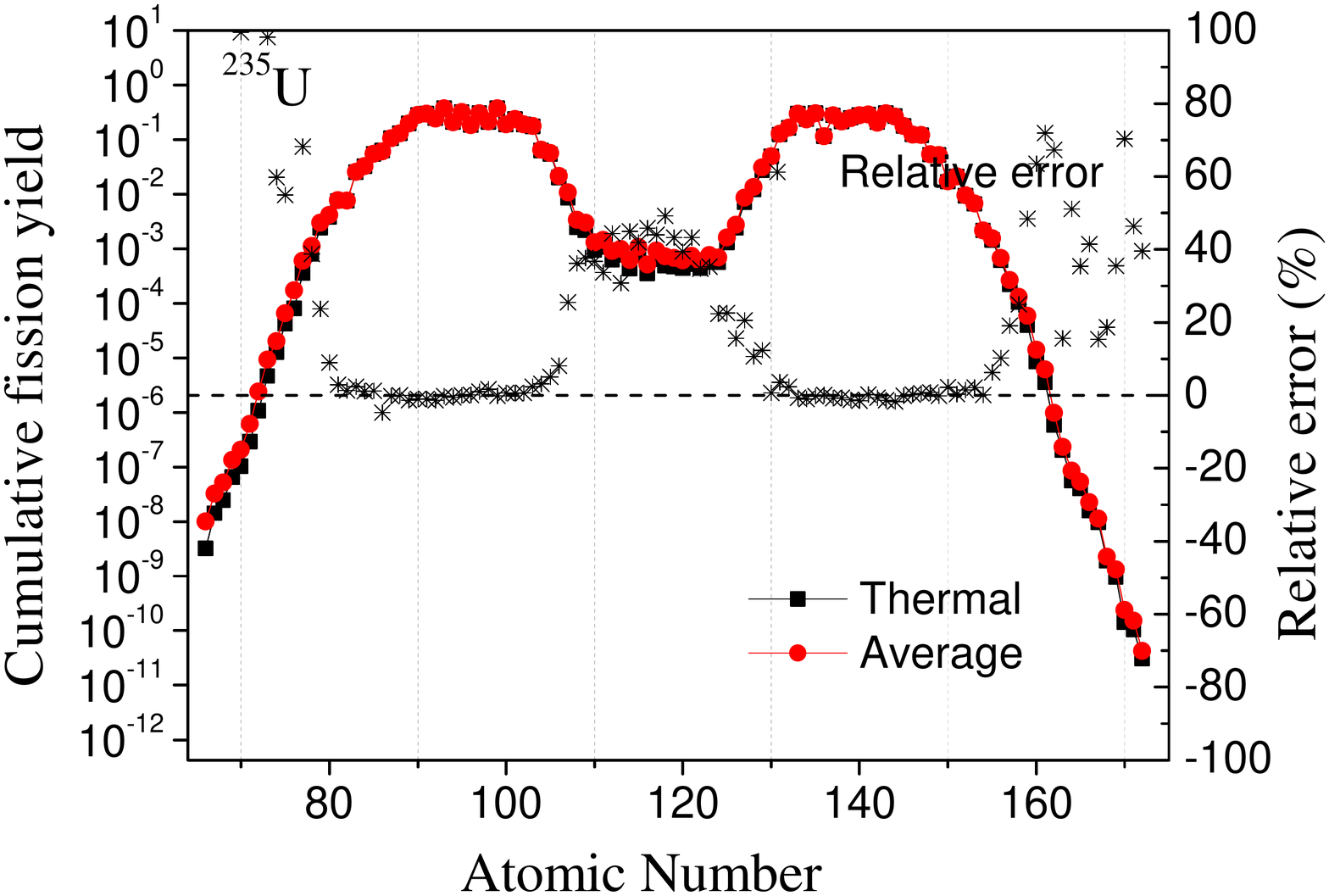}
\caption{Average fission yield distribution of $^{235}$U}
\label{u235yield}
\end{center}
\end{figure}

\begin{figure}
\begin{center}
\includegraphics[width=9cm]{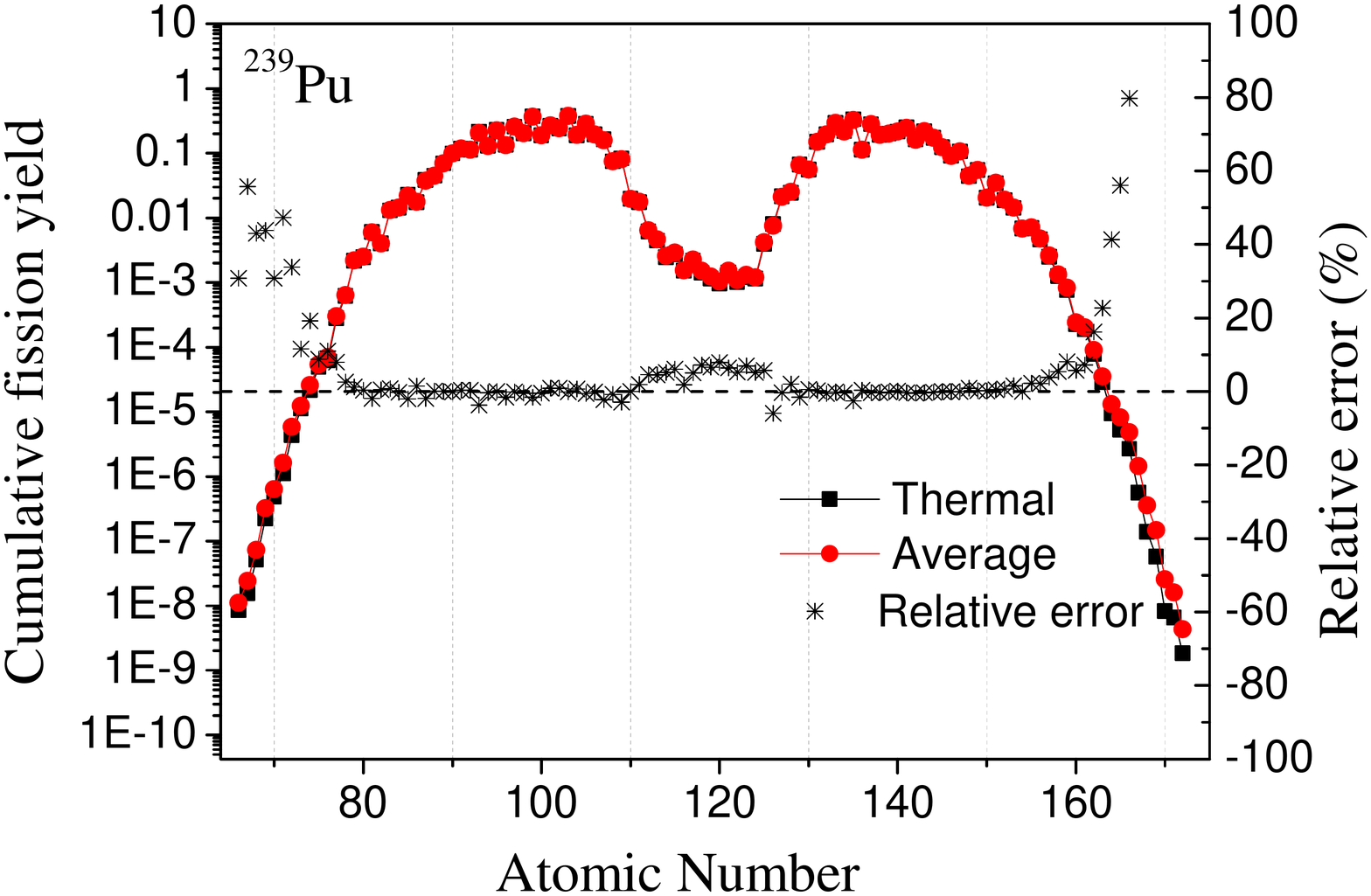}
\caption{Average fission yield distribution of $^{239}$Pu}
\label{pu239yield}
\end{center}
\end{figure}

\begin{figure}
\begin{center}
\includegraphics[width=9cm]{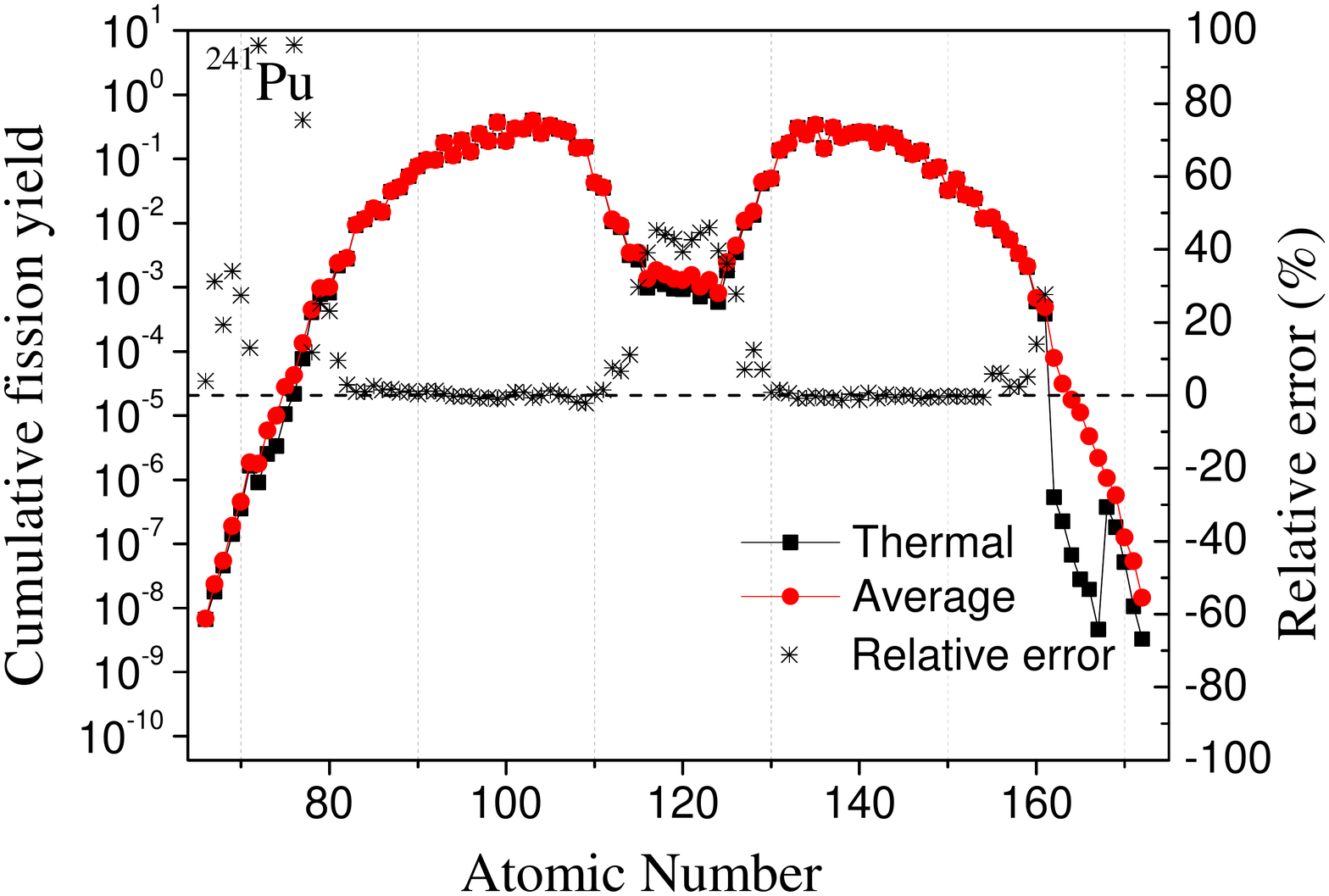}
\caption{Average fission yield distribution of $^{241}$Pu}
\label{pu241yield}
\end{center}
\end{figure}

To calculate the coefficients $C_{thermal}$ and $C_{fast}$ for each isotope, the fission fraction of the four isotopes with thermal and fast neutrons were calculated for a typical PWR using the SCIENCE code system\cite{Dayabay,spc1Dayabay,science}, which is used in the Daya Bay neutrino experiment reactor simulations. Fig.~\ref{thermalfastff} shows the thermal neutron and fast neutron fission fraction as a function of cycle burnup. The thermal neutron fission fraction is clearly larger than that of fast neutrons for $^{235}$U, $^{239}$Pu, and $^{241}$Pu. However, for $^{238}$U, the results are contrary. From Fig.~\ref{thermalfastff}, the average thermal and fast fission fractions and the thermal and fast coefficients were obtained (see Table \ref{cc}). The fast neutron fission yield is seen to have importance for $^{235}$U and $^{241}$Pu, and its contribution is about 23.18\% for $^{235}$U. Applying the thermal and fast neutron coefficients, the average fission-product yields of $^{235}$U, $^{239}$Pu, and $^{241}$Pu were obtained and they were compared with the thermal fission-product yields of Figs.~\ref{u235yield}, \ref{pu239yield}, and \ref{pu241yield}. There are in evidence large differences in the low, middle, and high atomic number regions. However, in these regions, the fission-product yields are always small. In this calculation, we did not take into account contributions from $^{238}$U because its fast neutron coefficient is 100\% and does not contribution to the bump because of the neutron spectrum. Muller's $^{238}$U spectrum was used when calculating the total antineutrino spectrum.

\begin{table}
\begin{center}
\caption{\label{cc}
Average fission fraction of the four isotopes (\%)}
\footnotesize
\begin{tabular*}{100mm}{c@{\extracolsep{\fill}}ccccc}
\toprule
              &$^{235}$U & $^{238}$U& $^{239}$Pu & $^{241}$Pu \\ \hline
 fast neutron & 13.4     & 7.5      & 2.84       & 0.91       \\ \hline
 thermal neutron & 44.38 & 0.0      & 26.32       & 4.46       \\ \hline
 total           & 57.78 & 7.5      & 29.17       &5.37        \\ \hline
 $C_{fast}$        & 23.18 & 100.0    & 9.75        & 16.89      \\ \hline
 $C_{thermal}$     & 76.82 & 0.0      & 90.25       & 83.11      \\
\bottomrule
\end{tabular*}
\end{center}
\end{table}

\section{Bump analysis of the neutron spectrum}
\label{bumpan}
\begin{figure}
\begin{center}
\includegraphics[width=9cm]{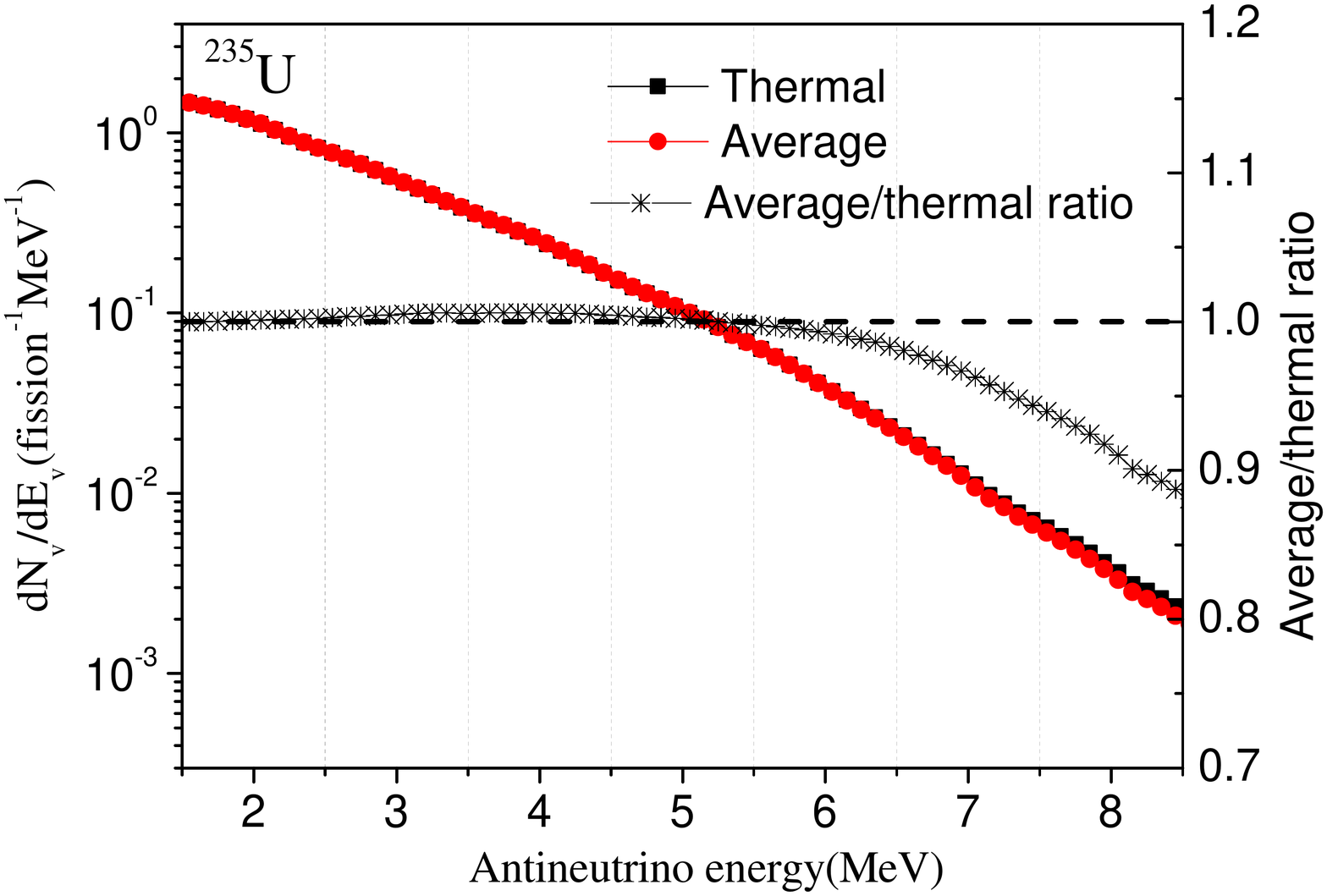}
\caption{ Antineutrino spectrum of $^{235}$U and the average-to-thermal ratio}
\label{u235spc}
\end{center}
\end{figure}

\begin{figure}
\begin{center}
\includegraphics[width=9cm]{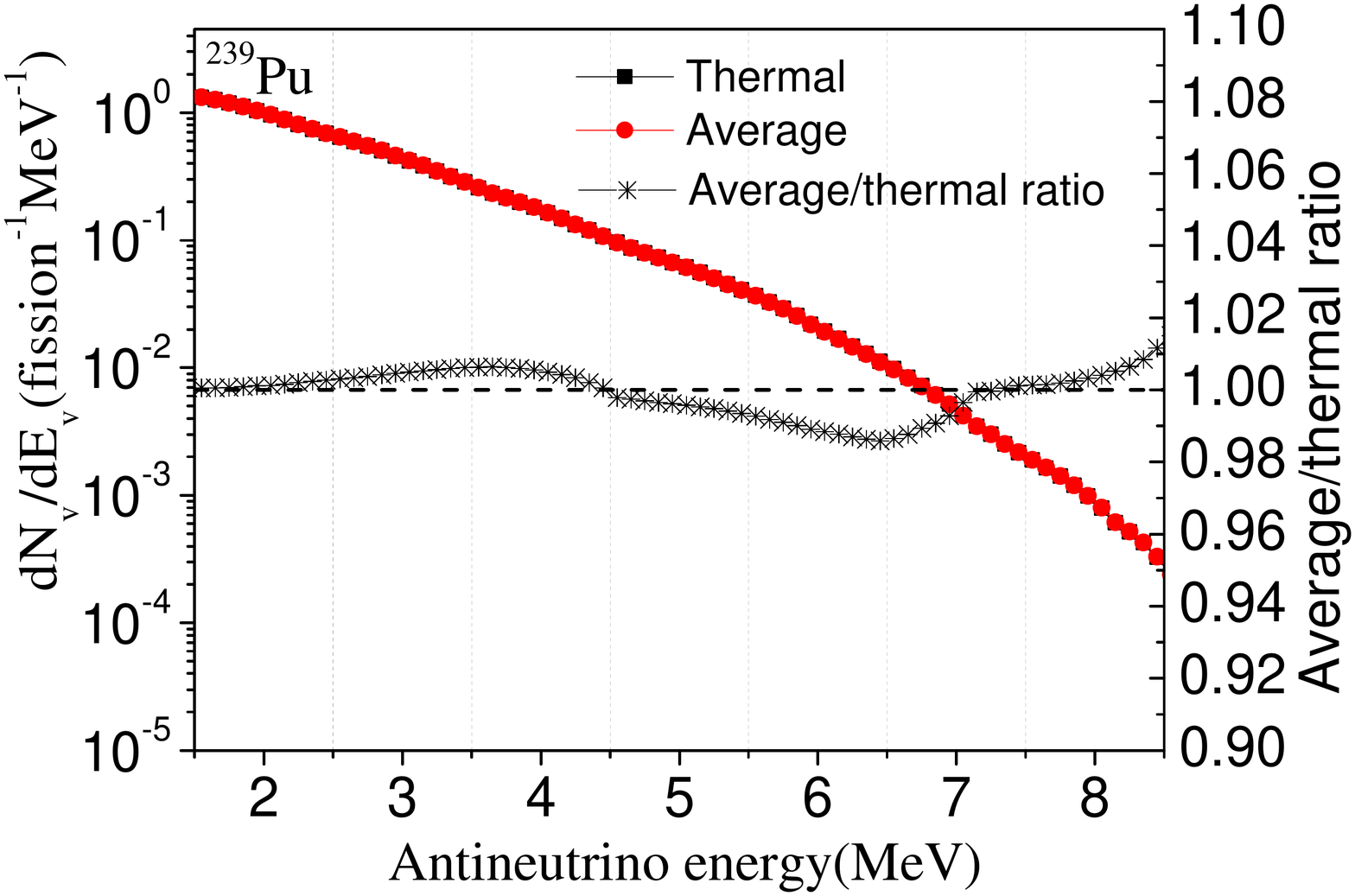}
\caption{Antineutrino spectrum of $^{239}$Pu and the average-to-thermal ratio}
\label{pu239spc}
\end{center}
\end{figure}

\begin{figure}
\begin{center}
\includegraphics[width=9cm]{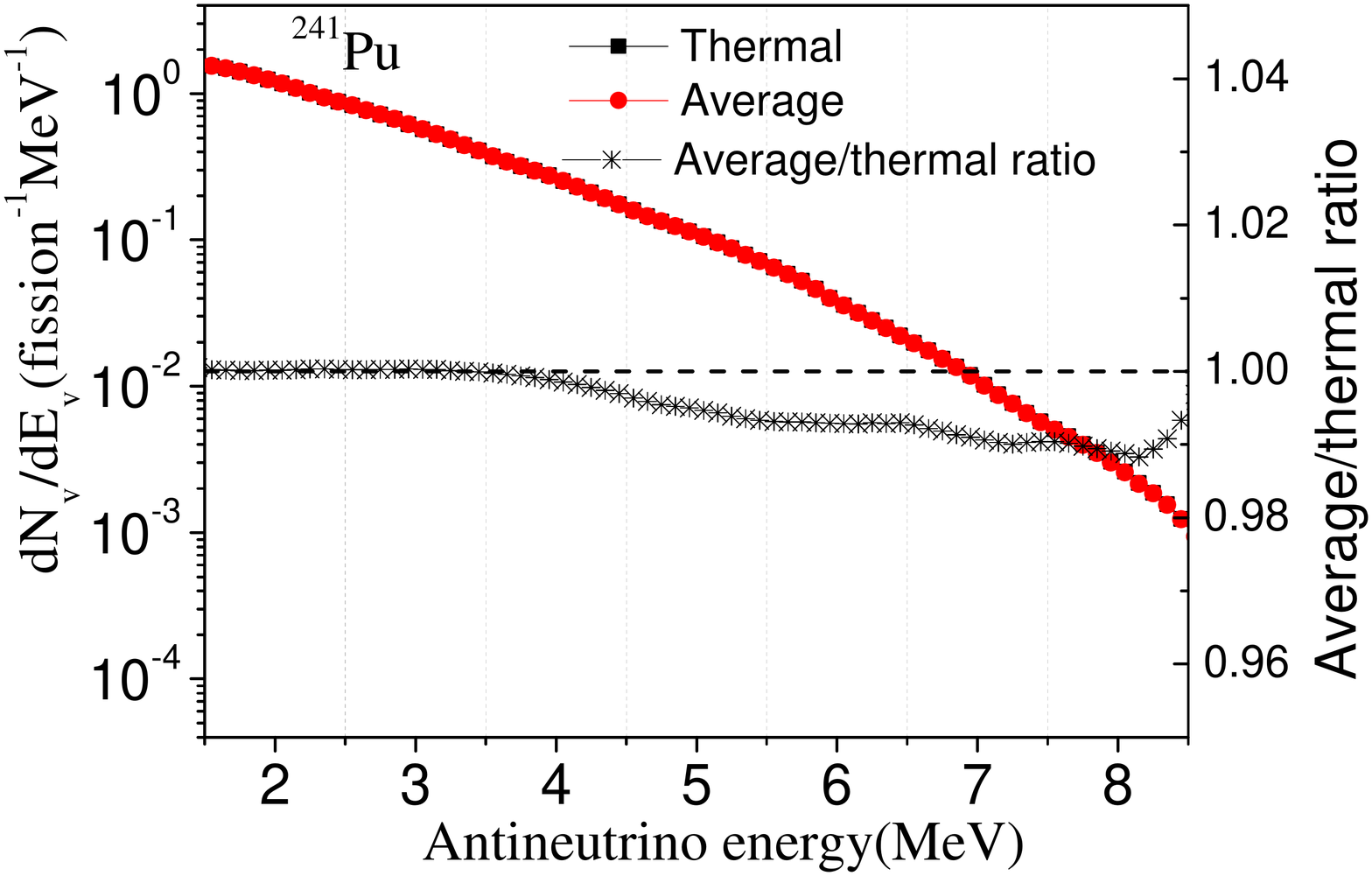}
\caption{Antineutrino spectrum of $^{241}$Pu and the average-to-thermal ratio}
\label{pu241spc}
\end{center}
\end{figure}

\begin{figure}
\begin{center}
\includegraphics[width=9cm]{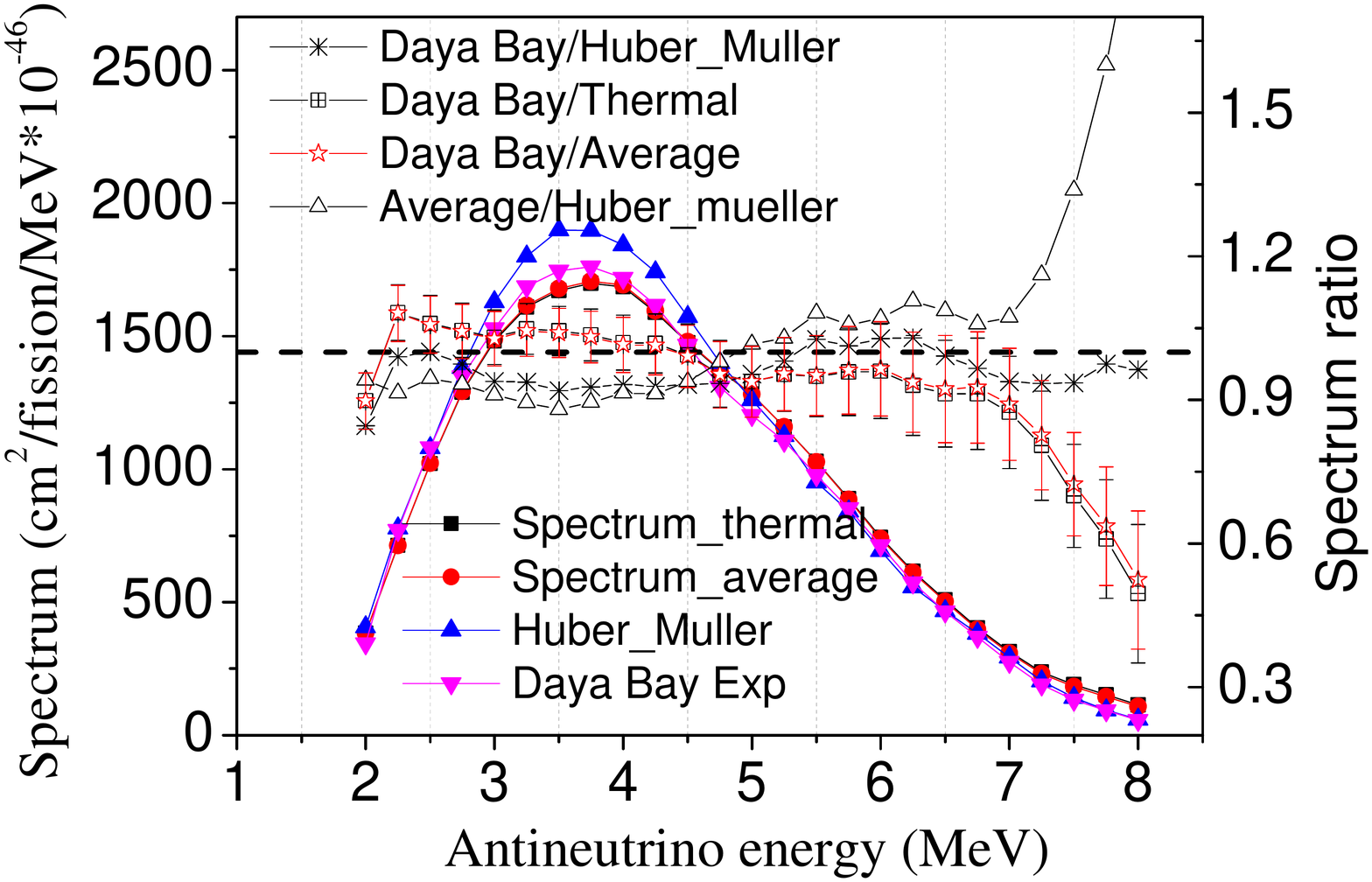}
\caption{Total antineutrino spectra and spectra obtained from the different models }
\label{comparison}
\end{center}
\end{figure}

Recent measurements of the positron energy spectra from $\bar{\nu}_{e}$ inverse beta decay ($\bar{\nu}_{e}+p\rightarrow e^{+}+n$) show an excess from 4 to 6~MeV, which corresponds to an excess of $\bar{\nu}_{e}$ spectra from 5 to 7~MeV because of the kinematic relationship $E_{\bar{\nu}_{e}}\sim E_{e^{+}} +0.8$~MeV. Using the average fission yield, the antineutrino spectra of $^{235}$U, $^{239}$Pu, and $^{241}$Pu were calculated and they were compared with that of the only-thermal neutron fission yield. The spectral ratio of the average fission yield to the thermal fission yield of $^{235}$U, $^{239}$Pu, and $^{241}$Pu are shown in Figs.~\ref{u235spc}, \ref{pu239spc}, and \ref{pu241spc}, respectively. There is no obvious excess from 5 to 7~MeV for these isotopes. However, the average-to-thermal ratio for $^{235}$U decreases with increasing antineutrino energy, the ratio being about 0.925 at energy 8.0~MeV. The spectra and the average-to-thermal ratio (Fig.~\ref{comparison}) show that the average-to-thermal ratio from the DayaBay data has no bump in the energy region 5--7~MeV. However, for the ratio obtained from the Huber--Muller model, a bump appears; these results are also found in \cite{hayes1}. Uncertainties in the average-to-thermal ratio for the Daya Bay data are only taken into account in the uncertainty for fission yields of each isotope, and the Monte Carlo method is applied in propagating the error of fission yields. The differences in the ratio between using thermal fast average fission yield and the only-thermal fission yield are small, and therefore, the neutron flux cannot be responsibility for the bump. Nevertheless, it will affect the distribution of the antineutrino spectrum, especially in the high energy region.


\section{Conclusion}
At present, it is not clear what physical processes give rise to the neutrino spectra bump. Much attention has been focused on the reactor antineutrino anomaly, which arose from improved calculations of the antineutrino spectra derived from nuclear databases. In this study, the question of whether the bump arises from the neutron spectrum was studied, and the \emph{ab initio} method was applied to calculated the antineutrino spectrum based on the updated nuclear database. To calculate the average fission yield, the fission rate was used as a weight function. The bump appears when comparing the results obtained with Huber--Mueller's model for both the thermal fast average fission yield and the only-thermal fission yield. These results were also found in a previous study. The differences between the average antineutrino spectrum and only-thermal antineutrino spectrum are small, which cannot be responsibility for the bump in the energy region from 5--7~MeV. However, the neutron flux does affect the total antineutrino spectrum distribution, especially in the high-energy region.

\section*{Acknowledgments}
The work was supported by National Natural Science Foundation of China (Grant Nos. 11390383, 11875128) and the Fundamental Research Funds for the Central Universities (Grant Nos. 2018ZD10, 2018MS044). We thank Richard Haase, Ph.D., from Liwen Bianji, Edanz Group China (www.liwenbianji.cn/ac), for editing the English text of a draft of this manuscript.





\end{document}